\documentclass[preprint,notoc]{JHEP3}
\usepackage{epsfig}

\def\to{\rightarrow}

\def\bi{\begin{itemize}}
\def\ei{\end{itemize}}

\def\c1p{C1^\prime}

\def\tst{\tilde t}
\def\ttau{\tilde \tau}

\def\tg{\tilde g}

\def\tq{\tilde q}

\def\tw{\widetilde W}
\def\tz{\widetilde Z}
\def\alt{\lesssim}
\def\agt{\gtrsim}
\def\be{\begin{equation}}  
\def\ee{\end{equation}}  
\def\bea{\begin{eqnarray}}  
\def\eea{\end{eqnarray}}  

\def\sps1ap{SPS1a$^\prime$}

\title{Neutralino dark matter in mSUGRA/CMSSM \\ 
with a 125~GeV light Higgs scalar
}
\author{Howard Baer$^{a}$, Vernon Barger$^b$ and Azar Mustafayev$^{c}$ \\
$^a$Dept.\ of Physics and Astronomy, University of Oklahoma, Norman, OK 73019, USA\\
$^b$Dept. of Physics, University of Wisconsin, Madison, WI 53706, USA\\
$^c$William I. Fine Theoretical Physics Institute, 
University of Minnesota, Minneapolis, MN 55455, USA\\
E-mail: \email{baer@nhn.ou.edu}, \email{barger@pheno.wisc.edu},
\email{mustafayev@physics.umn.edu}}

\preprint{\vbox{UMN--TH--3036/12, FTPI--MINN--12/09}}

\abstract{
The minimal supergravity (mSUGRA or CMSSM) model is an oft-used
framework for exhibiting the properties of neutralino (WIMP) cold
dark matter (CDM). 
However, the recent evidence from Atlas and CMS on
a light Higgs scalar with mass $m_h\simeq 125$~GeV highly constrains
the superparticle mass spectrum, which in turn constrains the 
neutralino annihilation mechanisms in the early universe. 
We find that stau and stop co-annihilation mechanisms -- already highly stressed 
by the latest Atlas/CMS results on SUSY searches -- are nearly eliminated if indeed 
the light Higgs scalar has mass $m_h\simeq 125$~GeV. Furthermore, 
neutralino annihilation via the $A$-resonance is essentially ruled out in mSUGRA
so that it is exceedingly difficult to generate thermally-produced 
neutralino-only dark matter at the measured abundance. 
The remaining possibility lies in the focus-point region which now moves 
out to $m_0\sim 10-20$~TeV range due to the required large trilinear 
soft SUSY breaking term $A_0$. 
The remaining HB/FP region is more fine-tuned than before owing to 
the typically large top squark masses.
We present updated direct and indirect detection rates for neutralino dark matter, 
and show that ton scale noble liquid detectors will either discover 
mixed higgsino CDM or essentially rule out thermally-produced neutralino-only CDM
in the mSUGRA model.
}  
\keywords{Supersymmetry
Phenomenology, Supersymmetric Standard Model, Large Hadron Collider, Dark Matter}

\begin{document}

\section{Introduction}
\label{sec:intro}

Particle physics models including weak scale supersymmetry (SUSY) are highly motivated in that
quadratic divergences arising from the scalar sector of the Standard Model (SM) are rendered
merely logarithmic under supersymmetrization, leading to a solution to
the gauge hierachy problem, and its associated need for fine-tuning~\cite{witten,wss}.
Indeed, the supersymmetrized version of the SM -- the Minimal Supersymmetric Standard Model (MSSM) --
receives some well-known support from experiment in that the precisely-measured weak scale
gauge couplings nearly unify to a point at scales $Q\simeq M_{GUT}\simeq 2\times 10^{16}$~GeV,
as anticipated for a SUSY GUT model~\cite{gauge}. In addition, while the SM allows for
a Higgs boson mass in the range $m_{H_{SM}}\sim 115-800$~GeV, the MSSM requires a light SM-like
Higgs scalar with mass $m_h\sim 115-135$~GeV~\cite{mh}. Recent Higgs search results from Atlas and CMS
indicate a compelling excess of events beyond background in several different channels which 
is consistent with a Higgs mass of around 125~GeV~\cite{atlas_h,cms_h}, 
falling squarely within the rather tight SUSY window! If indeed $m_h$ is verified to be in the
$\sim 125$~GeV range, then minimal AMSB~\cite{amsb} and minimal GMSB~\cite{gmsb} 
models require exceedingly heavy sparticle 
mass spectra, and may be excluded from a naturalness point of view (for plots, see {\it e.g.}
Ref.~\cite{djouadi} or the Appendix of this paper). Minimal gravity mediation models can still be
allowed, but require large, non-zero values of the trilinear soft breaking parameter $A_0$~\cite{h125}.

Further motivation for weak scale SUSY comes from astrophysics, in that the bulk of matter
in the universe is dark and cold  and may be ascribed  to a massive neutral non-relativistic particle, 
of which there are no candidates available in the SM. In contrast, in SUSY theories
the lightest neutralino $\tz_1$ is considered as a prototypical WIMP 
(weakly interacting massive particle) candidate~\cite{z1dm}. 
The so-called ``WIMP-miracle'' has been touted
as indirect evidence for neutralino dark matter in that thermal production followed by freeze-out
of a massive weakly interacting particle roughly requires that particle to have mass
around the weak scale~\cite{peskin,feng}. 

Most analyses of neutralino dark matter have adopted the paradigm minimal supergravity (mSUGRA~\cite{msugra,bbo}
or CMSSM~\cite{cmssm}) model to compare against experimental searches. The parameter space of the mSUGRA model
is comprised of
\be 
m_0,\ m_{1/2},\ A_0,\ \tan\beta,\ sign(\mu ) ,
\ee
where $m_0$ is the common GUT scale scalar mass, $m_{1/2}$ is the GUT scale gaugino mass, $A_0$ is
the GUT scale trilinear soft breaking term and $\tan\beta$ is the weak scale ratio of Higgs 
field vevs, and $\mu$ is the superpotential Higgs mass term whose magnitude (but not sign) is
determined by the conditions for appropriate electroweak symmetry breaking (EWSB).
The mass of the top quark also needs to be specified and we take it to be  
$m_t=173.3$~GeV in accord with Tevatron results~\cite{top}.

In mSUGRA, the WIMP miracle arguments really apply to the case of neutralino annihilation
into lepton pairs which occurs via $t$-channel slepton exchange, with slepton masses in the
$30-80$~GeV range~\cite{bb1}. This slepton mass range has long been excluded by 
direct slepton pair searches at LEP2~\cite{lep2slep}. During the post-LEP2 era, the measured abundance
of neutralino cold Dark Matter (CDM) can be found in the mSUGRA model in just several distinct regions of parameter
space, each characterized by an enhanced neutralino annihilation cross section. 
These include:
\bi
\item The stau co-annihilation region~\cite{stau} at low $m_0$ where $m_{\ttau_1}\simeq m_{\tz_1}$
such that neutralinos co-annihilate with staus which are thermally present in the 
early cosmic plasma.
\item The $A$-funnel region~\cite{Afunnel} at high $\tan\beta\sim 50$ where 
the mass of the CP-odd Higgs boson $A$ is suppressed such that
$2m_{\tz_1}\simeq m_A$ and $\tz_1$ pair annihlation is enhanced by the presence of a broad
$s$-channel $A$-resonance.
\item The hyperbolic branch/focus point region~\cite{hb_fp} (HB/FP) at large $m_0$ where
$|\mu |$ becomes small and the neutralino turns from nearly pure bino to mixed bino-higgsino. 
The large higgsino components of $\tz_1$ allow for enhanced neutralino
annihilation into $WW$, $ZZ$ or $Zh$.
\item The stop co-annihilation region~\cite{stop} at particular values of $A_0$ where
$m_{\tst_1}\simeq m_{\tz_1}$ so that neutralinos can co-annihilate with light stops.
\item The $h$-resonance region at low $m_{1/2}$ where $2m_{\tz_1}\simeq m_h$ so that
neutralinos have an enhanced annihilation rate via the narrow $s-$channel $h$ resonance~\cite{hfun}.
\ei
Most of the mSUGRA parameter space lies outside these regions. In the bulk of parameter space,
the $\tz_1$ is pure bino-like and annihilates mainly through suppressed $p$-wave channels.
The neutralino relic density $\Omega_{\tz_1}h^2$ tends to be of order 10-1000, or about 2-4 orders of 
magnitude above the measured abundance~\cite{bbs}. Thus, the bulk of mSUGRA parameter space
would be excluded under the assumption of thermally-produced neutralino-only (TPNO) CDM.

A further blow to the neutralino-only CDM scenario in mSUGRA was delivered by early LHC 
gluino and squark searches, 
which so far have seen no sign of SUSY particle production for analyses within the context of the
mSUGRA model. Based on searches using just $\sim 1$~fb$^{-1}$ of data~\cite{atlas,cms}, the
Atlas and CMS experiments exclude $m_{\tg}\alt 1$~TeV in the case of $m_{\tq}\sim m_{\tg}$,
and exclude $m_{\tg}\alt 700$~GeV in the case where $m_{\tq}\gg m_{\tg}$.
These negative search results have now excluded 
\begin{enumerate}
\item all of the $h$-resonance annihilation region,
\item the preferred (least fine-tuned) portions of the stau and stop co-annihilation region and
\item the preferred (lowest $m_0$ and $m_{1/2}$) portions of the $A$-resonance annihilation region.
\end{enumerate}
The HB/FP region has emerged relatively unscathed by the LHC gluino/squark searches, 
owing to the predicted very heavy squark masses.

A third blow to the neutralino-only scenario of SUSY CDM is associated with the recent 
announcement of evidence for a SM-like Higgs boson with mass $m_h\sim 125$~GeV; 
this topic forms the subject of this paper.

The goal of our paper is to outline consequences for thermally-produced neutralino-only  
CDM in the mSUGRA model if the 125 GeV Higgs signal turns out to be real. 
We find in Sec.~\ref{sec:Oh2} that all of the neutralino annihilation mechanisms in mSUGRA 
are highly stressed, making neutralino-only CDM in mSUGRA appear even more unlikely, 
although it is still not completely excluded. The last remaining hope lies in the
focus point region, which now moves out to much higher values of $m_0$ where very heavy
top squarks are disfavored by fine-tuning arguments.
In Sec.~\ref{sec:detect}, we assess the prospects for direct or indirect detection of
neutralino CDM in the remaining vestiges of parameter space. We find that the next
round of direct dark matter detection (DD) and indirection detection (IDD) experiments 
will likely either discover neutralino CDM or essentially  close the remaining possible 
parameter space regions, thus making a definitive test of the TPNO CDM hypothesis in mSUGRA.
In Sec.~\ref{sec:conclude} we present a summary and conclusions.

\section{Neutralino relic abundance when $m_h\simeq 125$~GeV}
\label{sec:Oh2}

In this section, we examine the implications of a 125~GeV Higgs boson $h$ for the
neutralino relic density $\Omega_{\tz_1}h^2$ in the mSUGRA model. Some of these implications have 
already been obtained in Ref.~\cite{hww,h125} (see also Ref.~\cite{djouadi,fms,others,egn,ellisolive}). 
Here, we first recount some highlights of the previous calculations. 
From a scan over mSUGRA parameter space it was found that, for $m_h=125\pm 1$~GeV
\bi
\item $m_0\agt 0.8$~TeV. This result rules out the low $m_{1/2}$ portion 
of the stau co-annihilation region, although viable stau co-annihilation points 
can still occur at very high $m_{1/2}$ values.
\item A value $m_A\agt 1$~TeV is required. 
In addition, for a scan over all $A_0$ and $\tan\beta$ values for $\mu >0$, 
it was found that the lower quadrant of the $m_0\ vs.\ m_{1/2}$ plane is excluded
with $m_{1/2}\alt 1$~TeV (for low $m_0$) and $m_0\alt 2$~TeV (for low $m_{1/2}$).
Comparing to displays of the $A$-resonance annihilation region in 
Ref's~\cite{csaba,chi2,bmpt} and \cite{Afunnel}, 
these results eliminate the lower  portion of this region.
\item Since the top squark co-annihilation region tends to occur
at distinct non-zero values of $|A_0|$, but at low $m_0$ and $m_{1/2}$~\cite{stop}, 
this excludes much (but not all)
of the stop co-annihilation region.
\item The range $|A_0 |\alt 1.8 m_0$ is excluded for scans in $m_0$ up to 5~TeV. 
If the scan is enlarged to $m_0$ up to 20~TeV, then the excluded range diminishes to
$|A_0|\alt 0.5 m_0$. We note here that the HB/FP region with low $\mu$
occurs at $m_0\sim 3-10$~TeV for $A_0\simeq 0$. However, as $|A_0|$ increases, the 
$X_t$ term in the renormalization group equations for $m_{H_u}^2$ increases, which means
$m_{H_u}^2$ is driven more sharply negative. Since $\mu^2\simeq -m_{H_u}^2$~\cite{nuhm1}
at the weak scale, this means the HB/FP regions is pushed for non-zero $A_0$ 
values out to much larger $m_0\sim 10-50$~TeV values. 
While HB/FP annihilation can survive, it does so at
such high $m_0$ values that large third generation squark masses can lead to rather 
severe electroweak fine-tuning.
\ei
In summary, recent LHC searches for gluinos and squarks, if combined with the 
requirement of $m_h\sim 125$~GeV, rule out most of the remaining regions of
mSUGRA parameter space where one can obtain $\Omega_{\tz_1}h^2\simeq 0.11$. However,
some regions do remain. In this Section, we characterize these remaining regions.

To do so, we implement a scan over the mSUGRA parameter space:
\bea
m_0&:& 0\to 5\ {\rm TeV}\ \ ({\rm blue\ points});\ \  m_0:\ 5\to 20\ {\rm TeV}\ \ ({\rm orange\ points}) ,\
\\
m_{1/2}&:& 0\to 2\ {\rm TeV},\\
A_0&:& -5m_0 \to \ +5m_0,\\
\tan\beta&:& 5 \to 55 .
\eea
We employed Isasugra 7.81~\cite{isasugra} to generate 30K random points in the above parameter space, 
requiring only the LEP2 chargino bound~\cite{lep2ino} $m_{\tw_1}>103.5$~GeV. 
The radiative electroweak symmetry breaking is maintained and the
lightest supersymmetric particle (LSP) is required to be the lightest neutralino $\tz_1$.
We scan separately over positive and negative $\mu$ values, noting here that $\mu >0$ is 
preferred if we do not wish to stray more than
$3\sigma$ away from the measured value of the muon anomalous magnetic moment, $(g-2)_\mu$~\cite{gm2}.
We calculate $\Omega_{\tz_1}h^2$ using the IsaReD~\cite{isared} subroutine of Isajet/Isasugra.
Each point which is accepted is required to have $m_h=125\pm 2$~GeV.

The results of our scan in the $\Omega_{\tz_1}h^2\ vs.\ m_{\tz_1}$ plane for $\mu >0$ are 
shown Fig.~\ref{fig:Oh2vmz1}{\it a}). 
The $3\sigma$ WMAP-allowed~\cite{wmap7} region $\Omega_{CDM}h^2=0.111 \pm 0.017$ is denoted by the green band. 
From frame {\it a}), which includes 8950 surviving points, we see that the bulk of points with $m_0<5$~TeV populate
$\Omega_{\tz_1}h^2\sim 1-100$. Only a single blue point lies within the WMAP-allowed
band, with $m_{\tz_1}\simeq 679$~GeV (a stau co-annihilation point), while another blue point lies well
below the WMAP band at $m_{\tz_1}\simeq 360$~GeV (a stop co-annihilation point). 
If we allow a scan of $m_0$ up to 20 TeV (orange points) then we
see that $\Omega_{\tz_1}h^2\sim 10-10^4$ becomes favored. However, now a band of 
orange points with $\Omega_{\tz_1}h^2 <0.11$ spans the low $m_{\tz_1}$ range, intersecting
the WMAP band around $m_{\tz_1}\sim 700$~GeV. These points correspond to mainly higgsino-like
neutralinos with a typical underabundance. In addition, we find 11 points out of a sample of 30K with
$\Omega_{\tz_1}h^2$ (nearly) within the WMAP-allowed band. Inspection of each of these points revealed
that ten had mixed bino-higgsino DM, and so an enhanced annihilation rate. The eleventh
point -- the blue point with $m_{\tz_1}\simeq 679$~GeV -- lies within the remnant of the stau co-annihlation region, with
$m_0=835.1$~GeV, $m_{1/2}= 1551.2$~GeV, $A_0=-2151.1$~GeV and $\tan\beta =30$. For this point, 
$m_{\tg}=3.3$~TeV, $m_{\tq}\simeq 3$~TeV with $m_{\ttau_1}=679.3$~GeV and $m_{\tz_1}=678.6$~GeV. The required $\ttau_1 -\tz_1$
mass gap has dropped below 1~GeV. Such tiny mass gaps are needed to enhance stau co-annihilation in cases where the
stau masses are very large. The tiny mass gaps are highly fine-tuned in relic density as well in EWSB\cite{eo_ft,bss}.
\FIGURE[tbh]{
\includegraphics[width=12cm,clip]{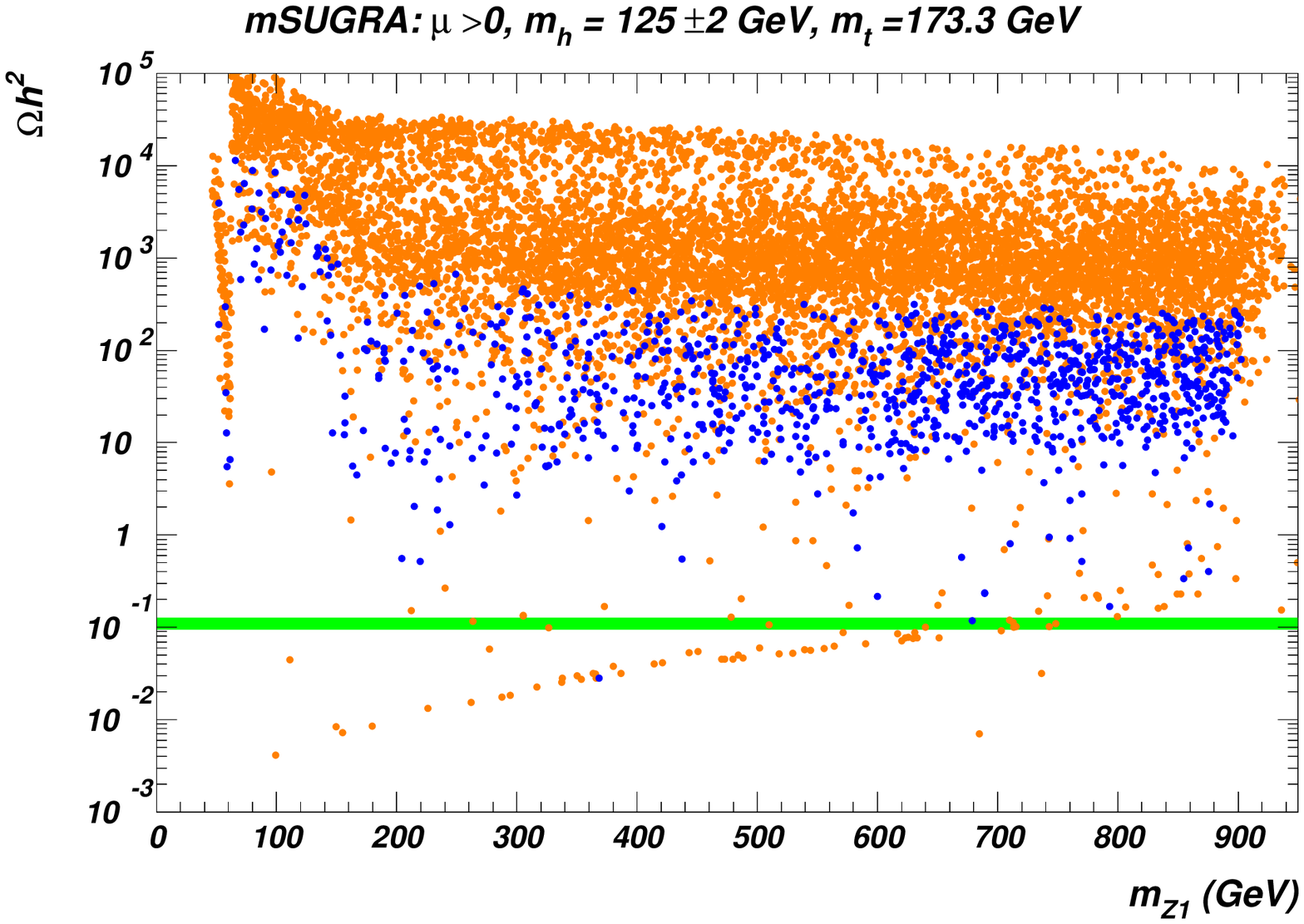}
\includegraphics[width=12cm,clip]{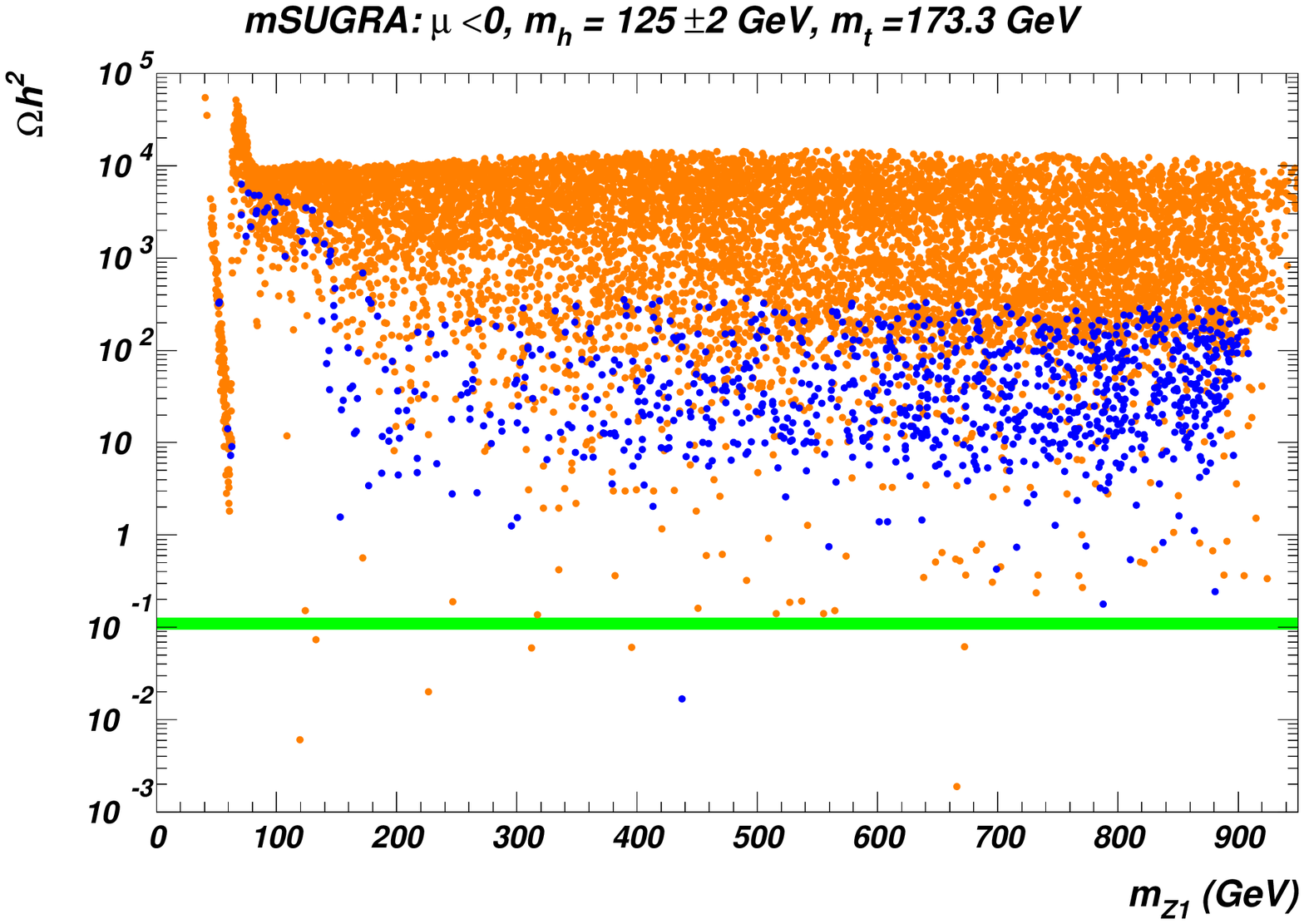}
\caption{
The neutralino relic density $\Omega_{\tz_1}h^2$ versus the neutralino mass $\ m_{\tz_1}$ 
from a scan over mSUGRA parameter space while requiring $m_h=125\pm 2$~GeV for {\it a}) $\mu >0$ and 
{\it b}) $\mu <0$. Blue and orange points correspond to $m_0 < 5$~TeV and $5{\rm TeV} < m_0< 20$~TeV,
respectively.
The shaded green horizontal bands represent the WMAP $3\sigma$ range.
}
\label{fig:Oh2vmz1}}

The results for $\mu <0$ are shown in  Fig.~\ref{fig:Oh2vmz1}{\it b}). The overall pattern
is similar to that of  Fig.~\ref{fig:Oh2vmz1}{\it a}): there is very little probability to generate
points with $\Omega_{\tz_1}h^2\sim 0.11$, and instead typically a vast overabundance of neutralino CDM is expected.
In the $\mu <0$ case, even fewer generated points lie in the $\Omega_{\tz_1}h^2\alt 0.11$ region.
This is in part because -- with fixed $m_{1/2}$ with $\Omega_{\tz_1}h^2\simeq 0.11$ -- 
the value of $m_{\tw_1}$ is significantly lighter in the $\mu <0$ case than for $\mu >0$. This means the
LEP2 chargino mass bound excludes a greater portion of the HB/FP region for $\mu <0$, and so the 
WMAP-allowed region is much narrower. Thus, there is a much smaller probability to generate
WMAP and LEP2 allowed points for negative $\mu$.

In our scans, we were unable to find any remaining $A$-funnel annihilation points.
To test what remains of the $A$-resonance annihilation region, we plot in Fig.~\ref{fig:Afun}
the value of $m_h$ versus the mass gap fraction $(m_A-2m_{\tz_1})/m_A$. 
For the $A$-resonance annihilation to be efficient, the $\tz_1$ mass should be within a few widths from $m_A/2$, which means
that $|m_A-2m_{\tz_1}|/m_A\alt 0.1$. 
From the plot, we see
that in this region essentially all models have $m_h<123$~GeV. Furthermore, those that are
WMAP-allowed with $\Omega_{\tz_1}h^2<0.128$ (green and red) all have $m_h<121.5$~GeV.
One must move out to a fraction $|m_A-2m_{\tz_1}|/m_A\sim 0.35$ to find a point with
$\Omega_{\tz_1}h^2<0.128$ and $m_h>123$~GeV. Given this scan, we conclude that the
$A$-resonance annihilation region is essentially excluded in mSUGRA by the Higgs mass requirement. In models such
as NUHM1~\cite{nuhm1} or NUHM2~\cite{nuhm2}, much lighter values of $m_A$ are allowed at any $\tan\beta$ 
value; in these cases, $A$-funnel annihilation can still be viable for TPNO CDM.
\FIGURE[tbh]{
\includegraphics[width=12cm,clip]{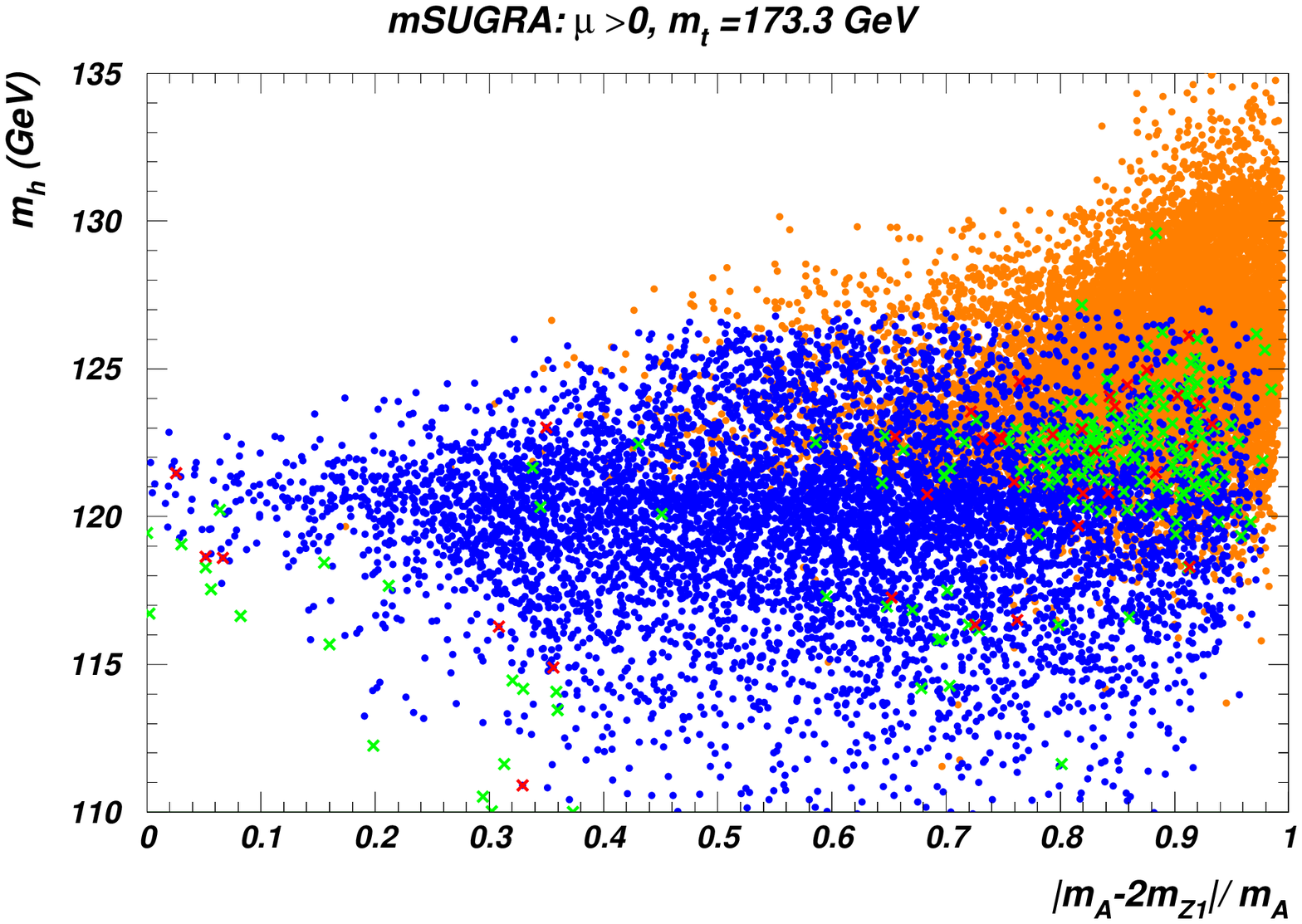}
\caption{
Light Higgs boson mass versus the neutralino-CP-odd Higgs boson mass gap
from a scan over mSUGRA 
parameter space.
Blue points denote $m_0 < 5$~TeV, while orange points allow $m_0$ values up to 20~TeV. 
Green and red crosses have the neutralino relic density $\Omega_{\tz_1}h^2<0.0941$ and 
$0.0941< \Omega_{\tz_1}h^2<0.1277$, respectively.
}
\label{fig:Afun}}
%

In Fig.~\ref{fig:Oh2}, we plot the number of models generated versus $\log(\Omega_{\tz_1}h^2)$
for the positive $\mu$ scan shown in Fig.~\ref{fig:Oh2vmz1}{\it a}). The WMAP-allowed
band is shown again in green. The plot can be interpreted as the probability to generate $\Omega_{\tz_1}h^2$ in
a particular bin, given a linear scan over mSUGRA parameters. From the plot, we see that there
is only a small probability to generate models with $\Omega_{\tz_1}h^2$ within the WMAP-allowed band: 
the vast bulk of models have $\Omega_{\tz_1}h^2\gg 0.1$ as shown by open black histogram.  
The $m_h=125\pm 2$~GeV requirement (blue histogram) significantly reduces the number of models, and especially 
in the WMAP-allowed region. There remains only a tiny probability to generate WMAP-compatible TPNO models, especially if one
requires not too heavy neutralinos $m_{\tz_1}<500$~GeV (as in Ref.~\cite{bbs}). 
We interpret the plot to imply
that the ``WIMP miracle'' is exceedingly difficult to realize in mSUGRA in the case
where the light Higgs scalar does indeed have a mass in the $125\pm 2$~GeV range.
\FIGURE[tbh]{
\includegraphics[width=12cm,clip]{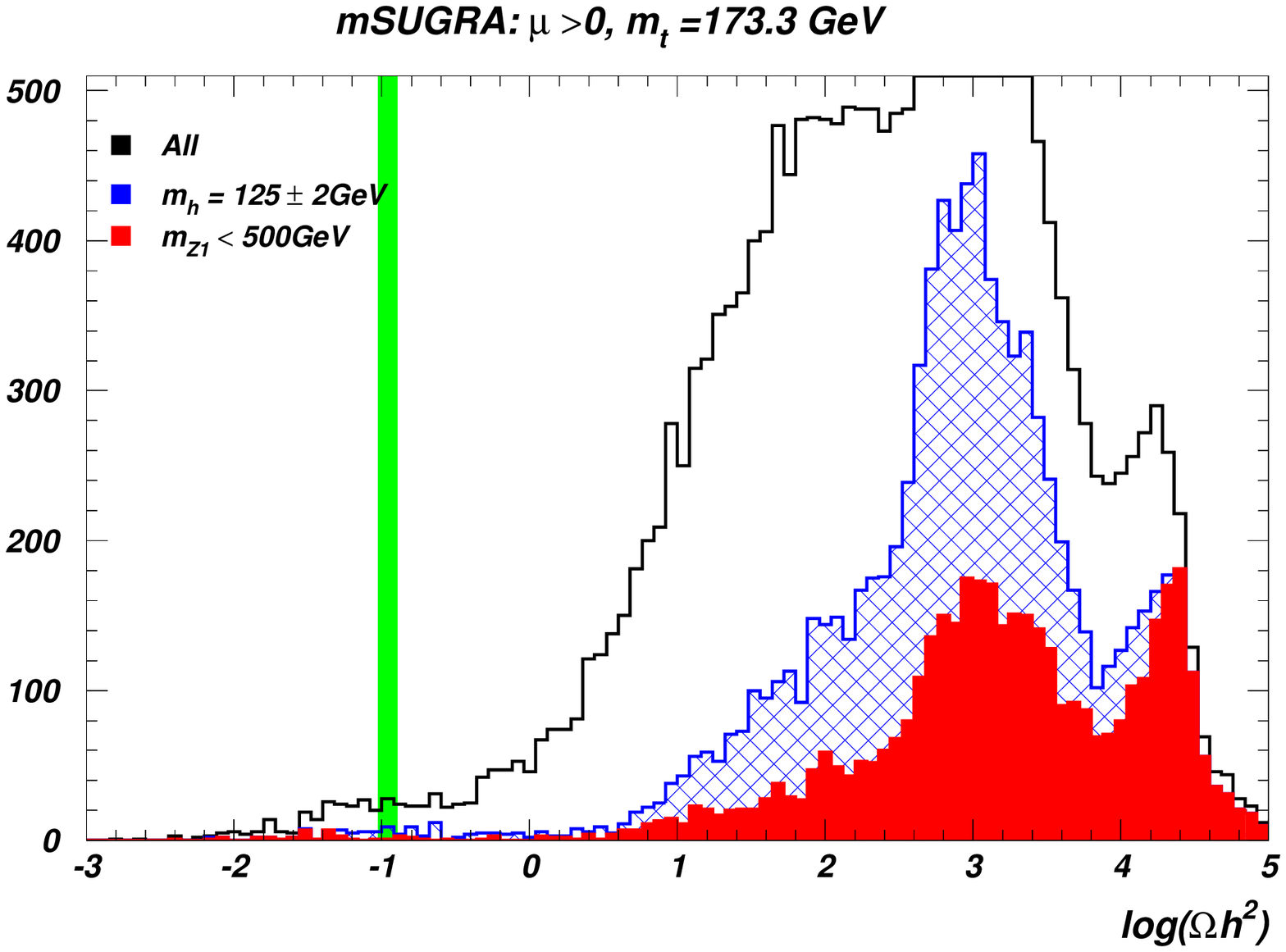}
\caption{
Distribution of neutralino relic density $\Omega_{\tz_1}h^2$ values in the
mSUGRA scan with $\mu >0$. 
The open black histogram represents all points satisfying the LEP2 chargino bound.
The hatched blue histogram requires $m_h=125\pm 2$~GeV,
while the filled red histogram additionally requires $m_{\tz_1}<500$~GeV.
The shaded green vertical band represents the WMAP $3\sigma$ range.
}
\label{fig:Oh2}}

While many of the WMAP-allowed points have mixed higgsino-like neutralinos with an underabundance
of apparent relic density, this doesn't necessarily mean these models would have an {\it actual}
underabundance. For instance, in models with additional multi-TeV scalar (moduli) fields, the moduli 
can undergo late decays at temperatures well below neutralino freeze-out, but above the temperature
where BBN starts. If the branching fraction to SUSY particles is significant, then
additional neutralinos can be injected into the cosmic plasma~\cite{mr,gg,kane}.
Similarly, if the strong $CP$ problem is solved by the Peccei-Quinn mechanism with a semi-invisible
axion, then thermal production of axinos or thermal/coherent production of saxions in the early universe can boost
the neutralino abundance~\cite{ckls,blrs,bls}. In these cases, the mixed higgsino-like neutralinos can still form the
bulk of CDM (the remainder being composed of perhaps axions), 
and may be searched for via direct or indirect dark matter detection experiments.

\section{Collider, direct and indirect neutralino detection rates}
\label{sec:detect}

\subsection{Detection at LHC}
\label{ssec:lhc}

While thermally-produced neutralino-only  CDM now seems rather improbable within the
mSUGRA model, it still remains a possibility. However, since it occurs mainly
with $m_0>5 $~TeV, this means squark pair production or gluino-squark associated production would be beyond LHC
reach, and LHC SUSY searches would mainly focus on gluino pair production signals.
In the case of decoupling squarks, the reach of LHC7 (or LHC8) for gluino pair production 
with 20~fb$^{-1}$ extends to $m_{\tg}\sim 1$~TeV~\cite{lhc7}, corresponding to $m_{\tz_1}\sim 150$~GeV.
The reach of LHC14 with 100~fb$^{-1}$ extends to $m_{\tg}\sim 1.7$~TeV~\cite{lhc14}, corresponding to 
$m_{\tz_1}\sim 250$~GeV. Thus, the green and red points with a mixed higgsino-like $\tz_1$ 
with $m_{\tz_1}\agt 300$~GeV would likely lie beyond LHC reach. A $pp$ collider with much higher
energy or luminosity would likely be needed to probe the remaining mixed higgsino branch.
In addition, it would still be possible to detect heavy neutralinos in direct or indirect
dark matter detection searches.

\subsection{Prospects for direct detection of neutralino CDM}
\label{ssec:dd}

To examine prospects for direct detection of neutralinos within the mSUGRA model, we plot
values of the spin-independent neutralino-proton cross section, $\sigma^{SI}(\tz_1 p )$, 
as a function of the neutralino mass, $m_{\tz_1}$, generated by IsaReS~\cite{bbbo} 
in Fig.~\ref{fig:z1p} from mSUGRA models with $m_h=125\pm 2$~GeV and $\mu >0$. 
The bulk of the points -- orange and blue points with $\Omega_{\tz_1}h^2$
well beyond WMAP limits -- give rise to very low direct detection rates. However, these models 
would be excluded under the hypothesis of thermally-produced neutralino-only CDM. 
The red points are consistent with the TPNO hypothesis, while the green points -- with a 
calculated underabundance of CDM -- would need their abundance augmented by other means, such as
late time scalar field or axino decay. We see that the allowed green and red points almost 
always lie below the current limits from the Xenon-100 experiment~\cite{xe100}, 
which means that Xenon-100 has only just recently moved to the upper edge of 
the remaining mSUGRA parameter space with a 125~GeV light Higgs.

The green/red swath occupies $m_{\tz_1}$ values from $\sim 100-800$~GeV, with
$\sigma^{SI}(\tz_1 p)$ values bounded from below by $\sim 10^{-9}$~pb. This is somewhat below
the usual HB/FP band~\cite{bbbo,wtn} which typically lies at $\sim 10^{-8}$~pb, due to the rather large 
squark masses $\sim 5-20$~TeV in the surviving models. 
The ultimate reach of Xenon 100~\cite{xe100}, which is comparable to the reach of LUX~\cite{lux},
can probe most of the remaining region, while Xenon 1ton~\cite{xe1t} can probe the remaining parameter space.
Thus, we expect Xenon 1ton or similarly sensitive ton scale noble liquid experiments to be able to 
either discover or exclude TPNO CDM in the mSUGRA model. 
We do note here that several WMAP-allowed points still
could survive Xenon 1ton scrutiny. The red point with $\left( m_{\tz_1},\sigma^{SI}(\tz_1p)\right) \sim \left( 
675\ {\rm GeV},10^{-11}\ {\rm pb}\right)$
is the previously mentioned last remaining vestige of stau co-annihilation, 
and with $m_{\tg}\sim m_{\tq}\sim 3$~TeV, lies at the reach limit of LHC14 with 100~fb$^{-1}$~\cite{lhc14}. 
The two green points with $\sigma^{SI}(\tz_1 p)\sim 10^{-14}$~pb
are remaining vestiges of the stop co-annihilation region. They have $m_{\tg}\simeq 2$ and $3.8$~TeV with
$m_{\tq}\simeq 4.6$ and 10.1~TeV respectively. 
The first green point with $m_{\tg}\simeq 2$~TeV value lies somewhat beyond, and the second green point well beyond, 
the 100~fb$^{-1}$ reach of LHC14.
\FIGURE[tbh]{
\includegraphics[width=12cm,clip]{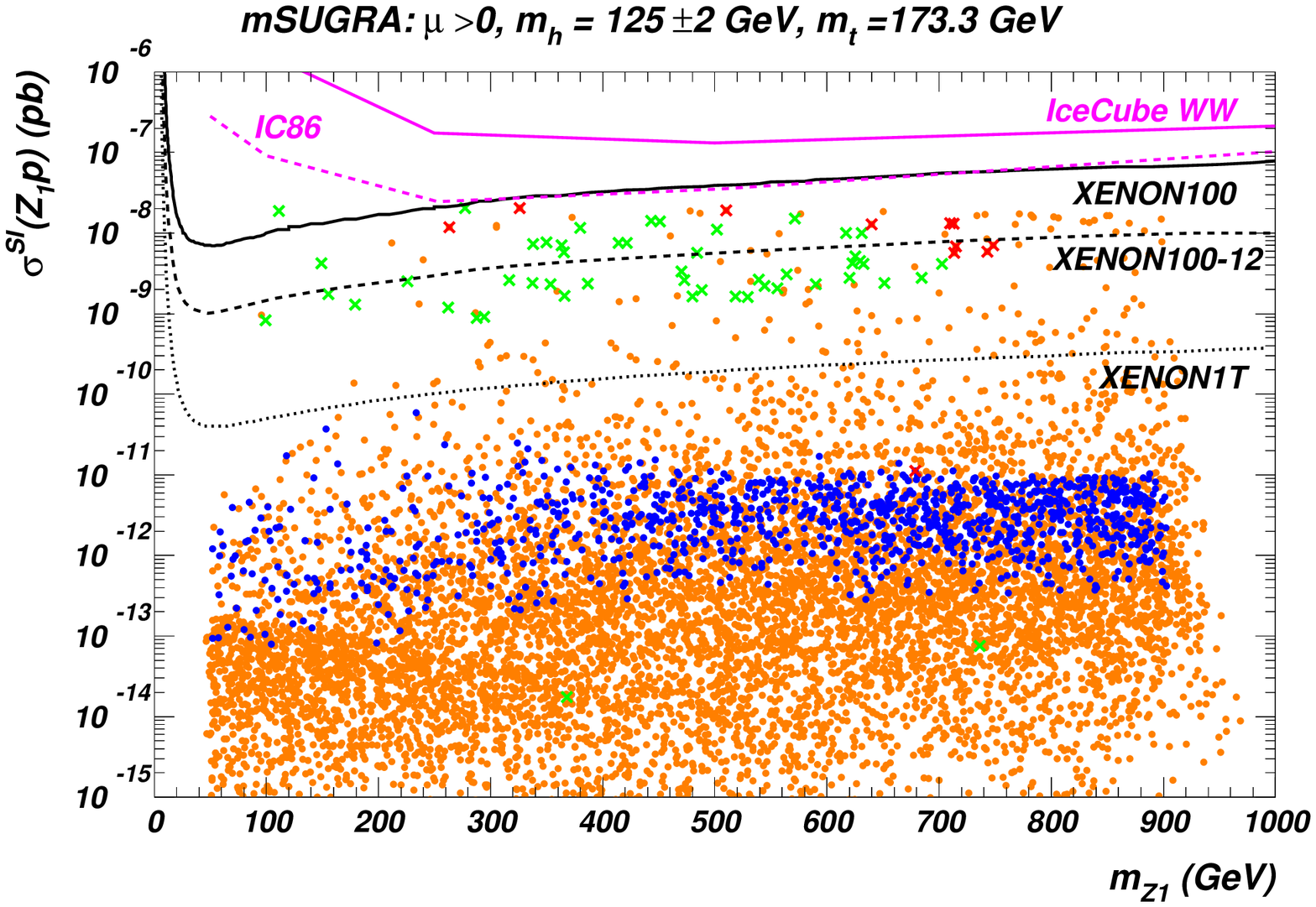}
\caption{
Spin-independent neutralino-proton elastic scattering cross section
 versus neutralino mass  
from a scan over mSUGRA parameter space while requiring $m_h=125\pm 2$~GeV.
Points color coding is the same as in Fig.~\ref{fig:Afun}.
We show the current and ultimate Xenon 100 exclusion limits~\cite{xe100} and the Xenon 1ton limit~\cite{xe1t}, for comparison.
We also show the curent and future limits from IceCube~\cite{icecube}.
}
\label{fig:z1p}}

In Fig.~\ref{fig:sd}, we plot the spin-dependent neutralino-proton elastic cross section
$\sigma^{SD}(\tz_1 p)$ versus $m_{\tz_1}$. 
The current direct detection limit from the COUPP experiment~\cite{coupp} reaches 
$\sigma^{SD}(\tz_1 p)\simeq 6\times 10^{-2}$~pb at $m_{\tz_1}\simeq 60$~GeV. 
We see that the WMAP-allowed red and green points
populate the $\sigma^{SD}(\tz_1 p)\sim 10^{-5}$~pb region, which is a few orders of magnitude below.
We also show the projected reach of the COUPP detector with 60~kg fiducial volume~\cite{coupp}; this experiment can cover
most but not all of the WMAP-allowed points.
\FIGURE[tbh]{
\includegraphics[width=12cm,clip]{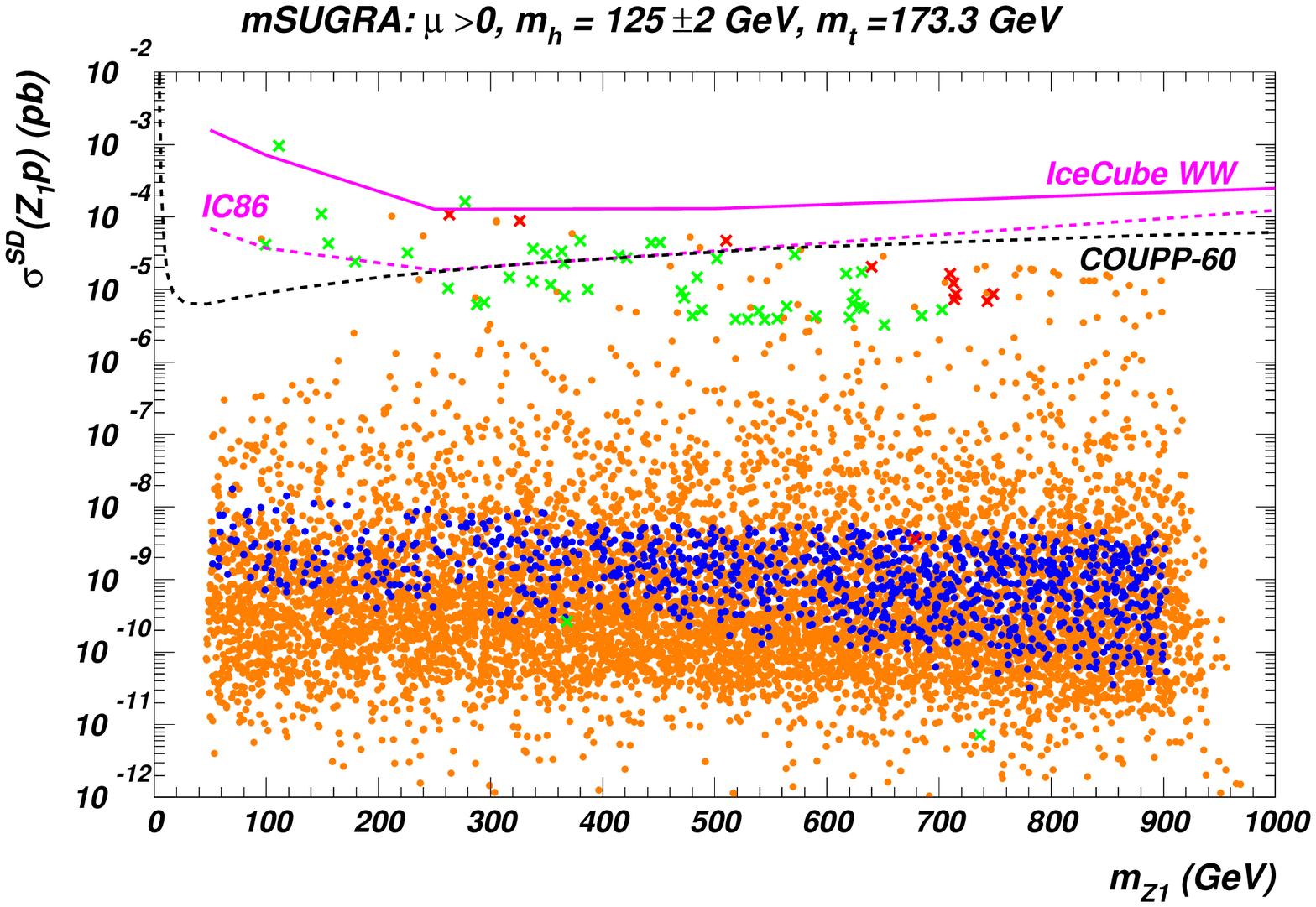}
\caption{
Spin-dependent neutralino-proton elastic scattering cross section
versus neutralino mass  
from a scan over mSUGRA parameter space while requiring $m_h=125\pm 2$~GeV.
Points color coding is the same as in Fig.~\ref{fig:Afun}.
We also show the curent and future limits from IceCube~\cite{icecube} and the future sensitivity 
of the COUPP experiment~\cite{coupp60}.
}
\label{fig:sd}}

\subsection{Prospects for indirect detection of neutralino CDM}
\label{ssec:idd}

The IceCube neutrino telescope located at the South Pole serves as an indirect
WIMP dark matter detector~\cite{fmw,bhhk,bbko}. 
As the Sun proceeds in its orbit about the galactic center, it would sweep up
WIMP particles which would ultimately collect in the solar core, whereupon they could annihilate 
into all allowed SM particle states. While most of these annihilation products would be absorbed in 
the solar medium, the high energy neutrinos from WIMP annihilation would escape, and could be
detected at IceCube via their conversion into high energy muons in the polar ice cap. 
The indirect detection rate of IceCube mainly depends on the SD and SI capture rate of
WIMPs by the Sun. There is also a weak dependence on the WIMP annihilation rate~\cite{jkg}.

In Figs.~\ref{fig:z1p} and \ref{fig:sd}, we also show limits from IceCube searches~\cite{icecube}. 
We only show the $WW$ channel, because almost all WMAP-allowed points come from the HB/FP region where $\tz_1$'s
dominantly annihilate into $W$ bosons, as explained in previous Sections. 
The current limits (solid magenta curves) lie at the upper 
edge of the allowed mSUGRA parameter space constrained by $m_h=125\pm 2$~GeV. 
The future IC86 curves show the completed IceCube detector (with 180 days exposure) will bite more deeply into this parameter space.

There is also the possibility to detect relic WIMPs from their annihilation within the galaxy or nearby
dwarf galaxies (or perhaps the galactic core), into antimatter ($e^+$, $\bar{p}$) or 
gamma rays ($\gamma$)~\cite{fmw,bbko}.
These detection rates mainly depend on the thermally averaged WIMP annihilation cross section
times velocity, evaluated at small velocities, $\langle\sigma v\rangle |_{v\to 0}$. This cross section, obtained from Isatools, is plotted
in Fig.~\ref{fig:sigv} versus $m_{\tz_1}$. For reference, we show the current limits from
the Fermi-LAT search~\cite{fermi} for $\gamma$s from WIMP annihilation into mainly $WW$, where the
$\gamma$s tend to arise mainly from $\pi^0\to\gamma\gamma$ from the $W$-boson decay and hadronization.
The WMAP-allowed green and red points populate the range 
$\langle\sigma v\rangle |_{v\to 0}\sim 10^{-26}-10^{-25}\ cm^3/sec$. 
The Fermi LAT search is already beginning to probe the allowed points, 
and would seemingly require $m_{\tz_1}\agt 200$~GeV. Further data samples and improved analyses
will push further into this parameter space.
\FIGURE[tbh]{
\includegraphics[width=12cm,clip]{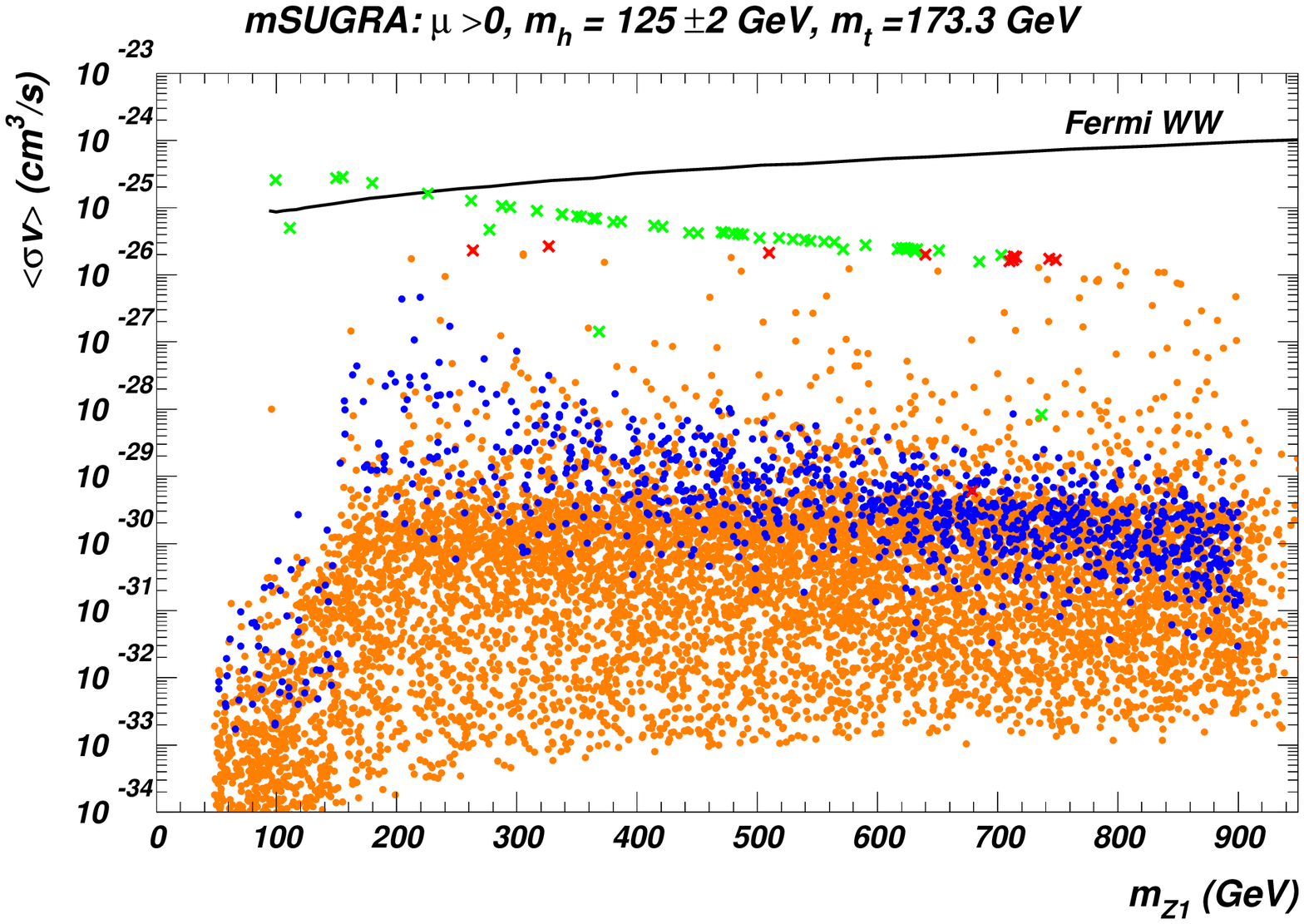}
\caption{
The thermally averaged WIMP annihilation cross section times velocity versus $m_{\tz_1}$ 
from a scan over mSUGRA parameter space while requiring $m_h=125\pm 2$~GeV.
Points color coding is the same as in Fig.~\ref{fig:Afun}.
We show the current limit from Fermi LAT~\cite{fermi}, for comparison.
}
\label{fig:sigv}}

\section{Conclusions}
\label{sec:conclude}

In the past, target dark matter search cross sections have been presented within the
context of the mSUGRA/CMSSM model under the assumption of the thermally-produced 
neutralino-only CDM hypothesis. For this model, neutralino CDM tends to be
grossly over-produced, and special enhancements to the neutralino annihlation cross section are
required to bring the model into WMAP compatibility. These enhancements include:
stau or stop co-annihilation, $A$ or $h$ resonance annihilation and mixed higgsino
annihilation, as occurs in the HB/FP region. Recent gluino and squark searches at LHC have
excluded the most lucrative co-annihilation and $A$-resonance annihilation regions, and
have completely excluded the $h$-resonance annihilation region. In this paper, we have also
examined the further constraint of requiring a light Higgs $m_h=125\pm 2$~GeV. Such a 
large value of $m_h$ requires large non-zero values of $A_0$, which pushes the HB/FP
region out to very high $m_0\sim 5-20$~TeV which may be in conflict with electroweak fine-tuning due to
large top squark masses. In addition, $m_h\sim 125$~GeV also requires
$m_0\agt 1$~TeV, so that squarks and sleptons are likely quite heavy. The combination of these
constraints rules out the $A$-resonance annihilation region and further squeezes
the stau and stop co-annihilation regions; the remaining co-annihilation
points, with large sparticle masses, require very narrow mass gaps indicating even more
fine-tuning of the neutralino relic density. 

While TPNO CDM in the mSUGRA/CMSSM model now seems even more unlikely than before the recent LHC results,
some possibilities do survive. The higgsino-like neutralino  branch tends to predict WIMP
detection rates that are within reach of the next round of direct and indirect DM detection 
experiments. If this branch is ruled out, then only a few extremely unlikely cases of
stop or stau co-annihilation will be left. These points will likely be beyond reach of
direct and indirect WIMP  detection experiments, and also likely beyond LHC sparticle
reach capabilities.

{\it Note added:} As this manuscript was completed, Ref. \cite{ellisolive} was released, 
performing similar studies. Their numerical results appear qualitatively similar to ours,
except their scan in $m_0$ is limited to values below 5 TeV. Thus, they do not obtain our
large $m_0>5$ TeV HB/FP results. 

\acknowledgments

We thank Francis Halzen for discussions. 
This work was supported in part by the U.S. Department of Energy under grants DE-FG02-04ER41305, 
DE-FG02-95ER40896 and DE-FG02-94ER-40823.

\appendix
\section{Appendix: Higgs mass in mAMSB and mGMSB}
\label{app:1}

In Ref.~\cite{djouadi}, the maximal value of $m_h$ was presented for various $\tan\beta$
values in the mGMSB model~\cite{gmsb} for $\Lambda$ values up to $10^3$~TeV, and in the mAMSB~\cite{amsb}
model for $m_{3/2}$ values up to $10^2$~TeV. Here, we show the predicted value of $m_h$ in these two models
from Isasugra for much higher parameter values until we reach the region where $m_h\simeq 125$~GeV.
In Fig.~\ref{fig:amsb}, we show the $m_h$ prediction in mAMSB versus $m_{3/2}$ for $\tan\beta =10$  and 45 
and for various $m_0$ choices. As can be seen, the value $m_h\simeq 125$~GeV is reached for 
$m_{3/2}\sim 600-800$~GeV.
At $m_{3/2}=600$ GeV and $\tan\beta =45$, $m_{\tg}\sim 10$~TeV, with $m_{\tst_1}\sim 9$~TeV and $\mu\sim 7$~TeV. 
This point certainly lies in the extreme electroweak fine-tuning region.
\FIGURE[tbh]{
\includegraphics[width=12cm,clip]{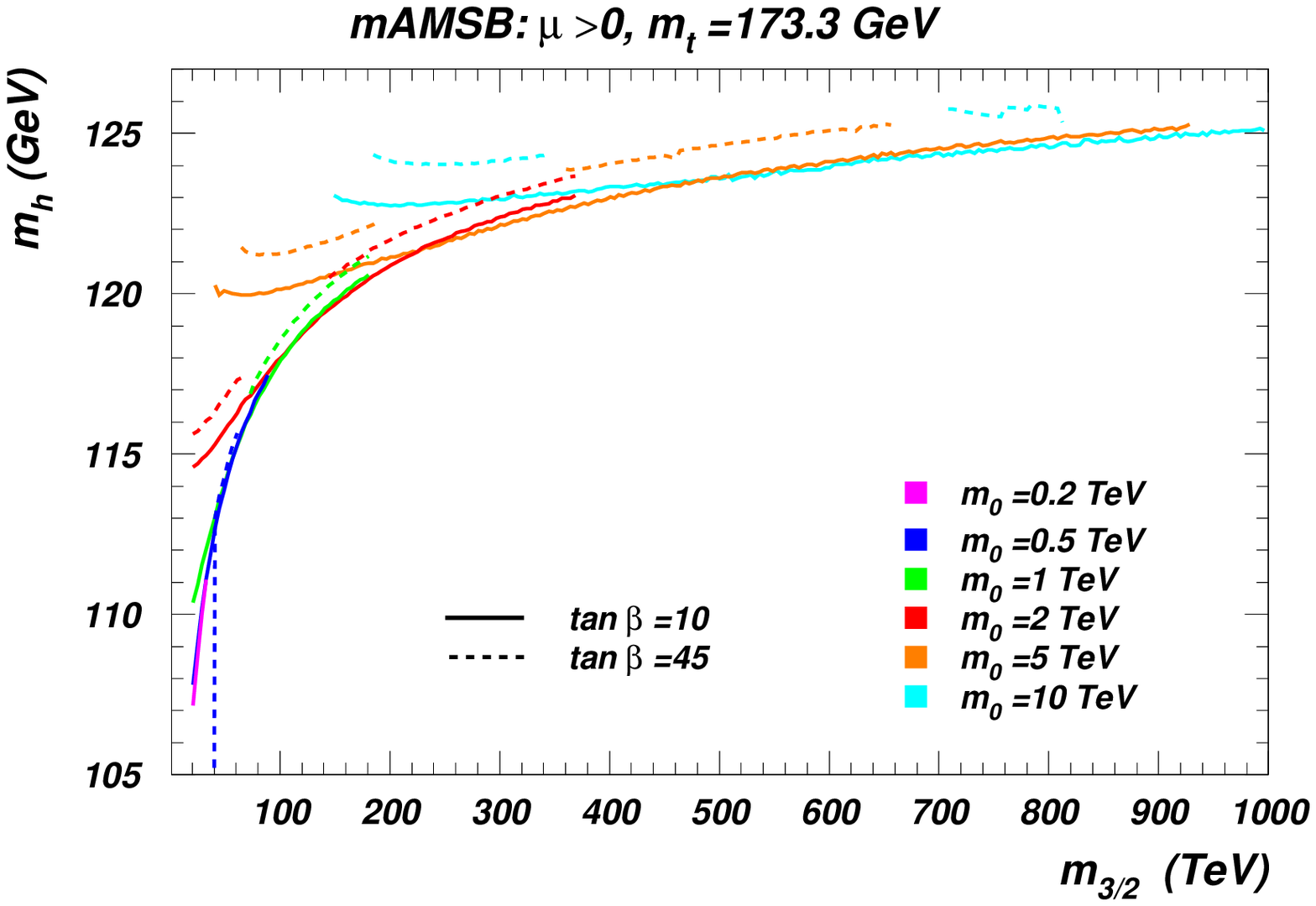}
\caption{
Light Higgs boson mass as function of the gravitino mass for various $m_0$ values
in minimal AMSB model for $\tan\beta =10$ and 45, and $\mu >0$.
}
\label{fig:amsb}}

In Fig.~\ref{fig:gmsb}, we plot the predicted value of $m_h$ from Isasugra for the mGMSB model
versus $\Lambda$ for $\tan\beta =10$ and 45,  messenger number $n_f=1$ and for messenger scale $M=2\Lambda$
and $M=10\Lambda$. In this model, where $A_0\sim 0$, we find $m_h\sim125$~GeV for
$\Lambda\sim 1500$~TeV (3000 TeV) for $\tan\beta =10$ (45). 
At the $\Lambda =1500$ TeV, $\tan\beta =45$ point, $m_{\tg}\sim 10$~TeV, with $m_{\tst_1}\sim 12$~TeV and
$\mu\sim 4$~TeV. This case is also in the regime of extreme electroweak fine-tuning.
\FIGURE[tbh]{
\includegraphics[width=12cm,clip]{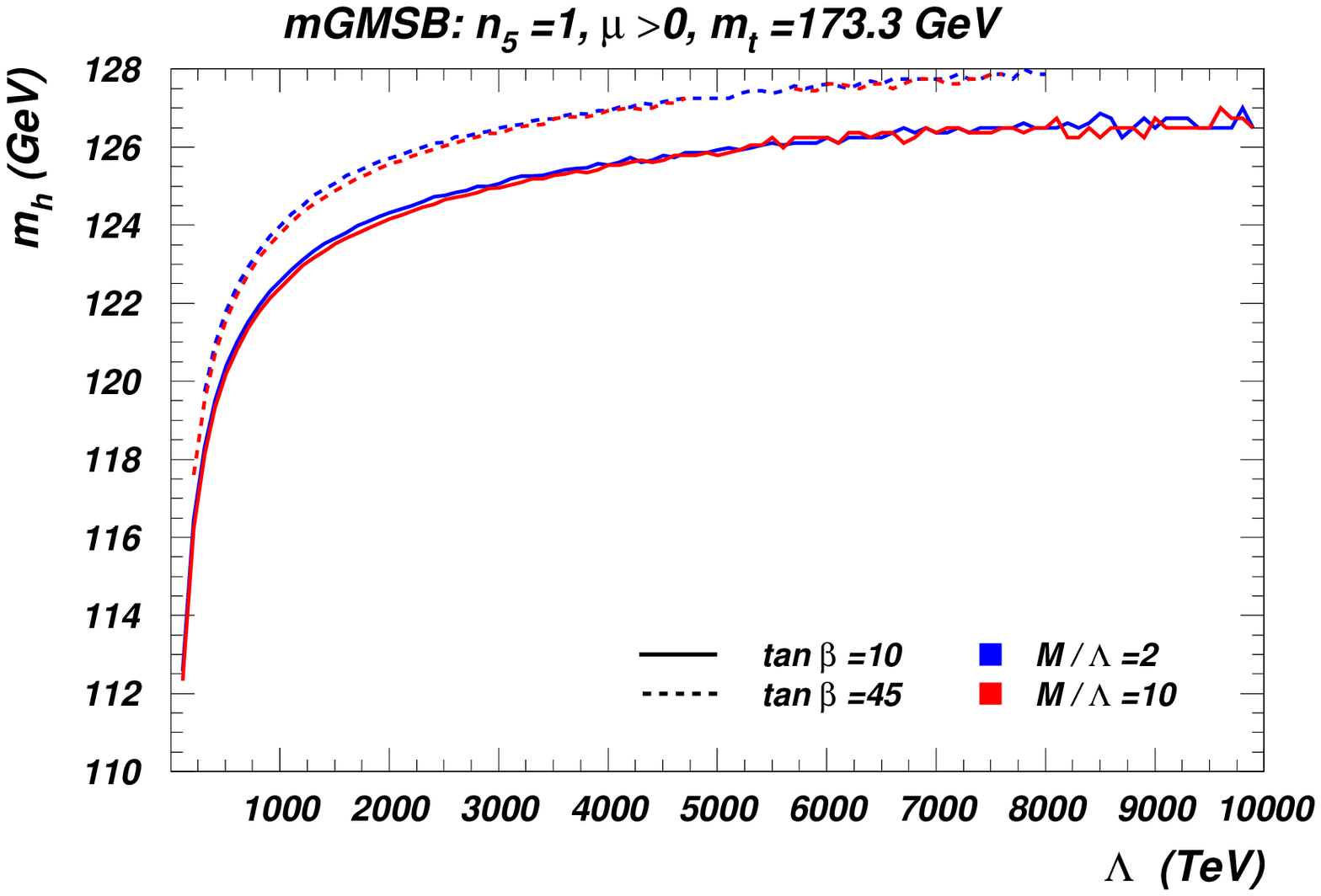}
\caption{
Light Higgs boson mass as function of the scale of the SUSY breaking for various values of the messenger mass
in minimal GMSB model for $\tan\beta =10$ and 45, and $\mu >0$.
}
\label{fig:gmsb}}
%

%

\end{document}